\documentclass[12pt]{article}
\begin{document}
\title{Gravitation and Electromagnetism\\
(Towards a Unified Description of Gravitation and Electromagnetism
Via ``Dark Energy")}
\author{G. S. Burra
}
\date{}
\maketitle
\begin{abstract}
Recently some hidden inconsistencies in high energy physics and
cosmology have been articulated by several scholars. If we follow
the usual description we get an unacceptably high cosmological
constant as was noticed by Weinberg and others some decades ago. In
this paper we point out that this is because of our continued
description in terms of the Planck scale. While this works for
gravitation, we have to consider a phase transition from the earlier
quantum foam to the Compton scale that describes electromagnetism,
and resolves the various problems. All this points towards the 
\textit{Theory Of Everything}.
\end{abstract}
\section{Introduction}
Recently Professor Harry Cliff of Cambridge University and LHC
Geneva has expressed his apprehension that physics is reaching a
dead end \cite{harry}. This is because of the un-acceptedly high
cosmological constant contradicting observation. We would like to
make a case that these contradictions arise because we stick to the
Planck scale. We will see how this scale leads to the real life
Compton
scale.\\
In 1997 the standard big bang cosmology had been sewed up. The
universe was decelerating, helped along by dark matter, which
comprised most of the universe. That year the author put forward an
alternative model, in which driven by what later came to be called
Dark Energy, the universe would be accelerating, though with a small
cosmological constant \cite{ijmpa1998,mg8}. This was observed the
very next year, in 1998, by S. Perlmutter, B. Schmidt and A. Riess
\cite{perl}.\\
In fact Anthony Leggett went on to say, "... It is of course clear
that your equation predicts an exponential (inflation-type)
expansion of the current universe, hence acceleration. And it would
have been nice if the Nobel committee had mentioned this, ..." and
"... I certainly do appreciate that you are one of the very few to
have recognized, on theoretical grounds, the possible need to
reintroduce a nonzero cosmological constant ahead of the supernova
experiment!" \cite{leggett}. He was referring to the fact that the
so called Dark Energy had been considered earlier but lead to the so
called cosmological constant problem: so large would be cosmological
constant, that the  universe would blow up almost as soon as it was
formed \cite{weinbergcc}.\\
In fact this form of Dark Energy, namely the well known Zero Point
Field had been considered almost a hundred years ago by Walter
Nernst who hypothesized that particles would be created out of the
vast pool of Dark Energy \cite{nernst}. However the ideas did not
lead to any fruitful conclusion. The whole point is that the Dark
Energy or Zero Point Field considered by Nernst, Zeldovich and
others differed subtly from that considered by the author in 1997.
We will investigate this point now and it would lead to some
interesting consequences such as providing a unified description of
gravitation and other interactions.
\section{The``New'' Dark Energy}
To put it in a nutshell the important difference between the two
Dark Energies is that the older variety was at the Planck scale,
whereas the 1997 model operated at the Compton scale.\\
To understand this we first observe that the Planck length is the
minimum length in the universe-- it describes a black hole of Planck
mass, $10^{-5} gms$, which forms an universe in itself as pointed
out by Rosen \cite{rosen}. It is at the junction of Classical Theory
and Quantum Theory. A rough way to see this is:
\begin{equation}
\frac{2Gm}{c^2} \sim \frac{\hbar}{mc}\label{A}
\end{equation}
The left side is the radius of a Classical black hole while the
right side is the Compton length, both these for the mass $m \sim
10^{-5}$. In fact to begin with the universe would be, what Wheeler
called the Quantum Foam, an ocean of Planck sized masses created
presumably by the Big Bang \cite{wheeler}. We will now see that such
a conglomoration will lead by a phase transition to the Compton
scale.
\section{The Phase Transition}
We start with space points $x$ in this Quantum Foam which in the
above sense represent Planck black holes. Let us first define a
complete set of base states by the subscript $\imath \quad
\mbox{and}\quad U (t_2,t_1)$ the time elapse operator that denotes
the passage of time between instants $t_1$ and $t_2$, $t_2$ greater
than $t_1$. We denote by, $C_\imath (t) \equiv < \imath |\psi (t)
>$, the amplitude for the state $|\psi (t) >$ to be in the state $|
\imath >$ at time $t,$ and \cite{ijpap,cu}
$$< \imath |U|j > \equiv U_{\imath j}, U_{\imath j}(t + \Delta t,t) \equiv
\delta_{\imath j} - \frac{\imath}{\hbar} H_{\imath j}(t)\Delta t.$$
We can now deduce from the super position of states principle that,
\begin{equation}
C_\imath (t - \Delta t) - C_\imath (t + \Delta t) = \sum_{j}
\left[\delta_{\imath j} - \frac{\imath}{\hbar} H_{\imath
j}(t)\right] C_\imath (t)\label{2ge}
\end{equation}
where the matrix $H_{\imath j}(t)$ is identified with the
Hamiltonian operator.\\
Equation (\ref{2ge}) leads in the limit to
\begin{equation}
\imath \hbar \frac{\partial \psi}{\partial t} = \frac{-\hbar^2}{2m'}
\frac{\partial^2 \psi}{\partial x^2} + \int \psi^* (x')\psi (x) \psi
(x')U(x')dx',\label{3eea8}
\end{equation}
In the above $U(x') = 1 \, \mbox{for} \, x'$ in a $\delta$ interval,
a small interval around this point and $= 0$ outside. The equation
(\ref{3eea8}) is a generalization of a two state equation made
famous by Feynman a long time ago (Cf.refs.\cite{feynman,cu,uof}).\\
We remember that considering the continuum,
$$H (x,x') = <\psi (x) | \psi (x')>$$
In (\ref{3eea8}), $\psi(x)$ is the probability amplitude of a
particle or point of the Quantum foam being at the point $x$ and the
integral is over a small $\delta$ region. It is easy to see that
(\ref{3eea8}) leads to the Landau-Ginsburg equation \cite{bgs2003}
\begin{equation}
- \frac{\hbar^2}{2m} \nabla^2 \psi + \beta |\psi |^2 \psi = - \alpha
\psi\label{3eea7}
\end{equation}
where we have generalized to the three dimensional case. The
similarity of (\ref{3eea8}) with (\ref{3eea7}) need not be
surprising considering also that near critical points, due to
universality, diverse phenomena like magnetism or fluids share
similar mathematical equations. Equation (\ref{3eea8}) was shown to
lead to the Schr\"{o}dinger equation with the particle acquiring a
mass (Cf.also ref.\cite{bgsnlw}). In the Landau-Ginsburg case there
is a coherence length which is given by
\begin{equation}
\xi = \frac{h \nu_F} {\Delta}\label{3eea9}
\end{equation}
where $\Delta$ is the energy $mc^2$ where $m$ is the mass of the
particle in the $\delta$ interval and $\nu_F$ is $c$. We can easily
show that in this case the coherence length reduces to the Compton
wavelength,
remembering that $\nu_F$ is $c$ in our case \cite{bgsnlw,bgsjsp}.\\
It may be mentioned that Beck and Meckay \cite{beck} also reached a
similar conclusion via the Landau-Ginsburg transition, though it is
not clear what their threshold of $1.7 Tz$ represents. Moreover,
Wigner and Salecker \cite{wigner} demonstrated that from the point
of view of a measurement, anything within the Compton scale is
unphysical. On the other hand the phase transition to the Compton
scale leads to the scale of electromagnetism. Whereas, the Planck
scale is associated with gravitation. We will see this now.
\section{Gravitation}
Cercignani \cite{cer} had used Quantum oscillations, though just
before the dark energy era -- these were the usual earlier Zero
Point oscillations. Invoking gravitation, what he proved was, in his
own words, "Because of the equivalence of mass and energy, we can
estimate that this (i.e. chaotic oscillations) will occur when the
former will be of the order of
\begin{equation}
G [\hbar \omega )c^{-2}]^2 [\omega^{-1}c]^{-1} = G\hbar^2 \omega^3
c^{-5},\label{B}
\end{equation}
where $G$ is the constant of gravitational attraction and we have
used as distance the wavelength. This must be less than the typical
electromagnetic energy $\hbar \omega$. Hence $\omega$ must be less
than
\begin{equation}
(G\hbar) ^{-1/2} \cdot c^{5/2}\label{C}
\end{equation}
which gives a gravitational cut off for the frequency in the
zero-point energy." In other words he deduced that there has to be a
maximum frequency of oscillators given by
\begin{equation}
G\hbar \omega^2_{max} = c^5\label{23}
\end{equation}
for the very existence of coherent oscillations (and so a coherent
universe). We would like to point out that if we use the above in
equation (\ref{23}) we get the well known relation
\begin{equation}
Gm^2_P \approx \hbar c\label{24}
\end{equation}
which shows that at the Planck scale the gravitational and
electromagnetic strengths are of the same order. This is not
surprising because it was the very basis of Cercignani's derivation
-- if indeed the gravitational energy is greater than that given in
(\ref{24}), that is greater than the electromagnetic energy, then
the Zero Point oscillators, would become chaotic and incoherent --
there would be no physics.\\
However all this refers to a classical description because we are
working here, at the Planck scale. In fact (\ref{23}) and (\ref{24})
can be alternatively deduced, considering these Planck oscillations
as phonons (Cf.ref.\cite{uof}). So the picture that emerges is the
following. Till frequencies with a cut off at the Planck length, a
classical description of the Zero Point energy is valid. This is the
domain of gravitation. Beyond that up to the Compton scale we have a
Quantum Mechanical description  and this leads to electromagnetism
and other interactions. In any case, its the Zero Point Field or
Dark Energy all the way.\\ \\
\noindent {\large {\bf Remark}\\ \\
We finally make the following
remark: The Landau-Ginzburg equation (\ref{3eea7}) comes from the
Lagrangian
\begin{equation}
{\bf L}_{GL} = L_{0} + a|\psi|^{2} + \frac{b}{2}|\psi|^{4} +
\frac{1}{2m}|\frac{\hbar}{i}\nabla - e{\bf A}\psi|^{2} + \frac{{\bf
B}^{2}}{2\mu_{0}}\label{A}
\end{equation}
So if we now construct the Lagrangian
\begin{equation}
{\bf L}_{U} = \alpha(t) {\bf L}_{GL} + \beta(t) {\bf
L}_{SM}\label{B}
\end{equation}
where $\alpha(t)$ and $\beta(t)$ are suitable coefficients, and
${\bf L}_{SM}$ is standard model Lagrangian, then we can see that
(\ref{B}) is a Lagrangian for all the four interactions in the light
of the above and because the standard model Lagrangian gives the
other three interactions. This is an approach towards grand
unification.
\section{A different Approach-- bi spinorial space}

The ultimate unification of gravitation and electromagnetism  eluded 
Einstein, 
while for a century physicists  have been trying to achieve this goal. 
Finally,  the great Pauli declared do not try to combine what Nature had 
meant 
to be kept separate. There were attempts over the decades by several 
physicists, for example,Hermann Weyl tried this unification.  Einstein, on 
the 
other hand  rejected it, on the grounds that (quite rightly),  gravitation 
was 
a force which was artificially inserted without any unification.The author 
has 
made an attempt at this unification, and this is described in his book 
``Thermodynamic Universe", (World Scientific) \cite{BGSTU-08} and ``The 
Universe of 
Fluctations" (Springer) \cite{uof} and other papers \cite{BGSnc-01}. Here the 
author 
brings out the fact that every point in space is bi spinnorial as will 
be 
seen below.
\section{Bi-Spinors}
It is a well-known fact that the Dirac  $4\times 4$ matrices,  are really 
made 
up of two  two-spinors.
Let us consider the usual spacetime,
in which the Dirac wave function is given by
$$\psi = \left(\begin{array}{ll}
	\chi \\ \Theta
\end{array}\right),$$
where $\chi$ and $\Theta$ are two component spinors. It is
known that under reflection, while the so called positive energy
spinor $\Theta$ behaves normally, on the contrary $\chi \to -\chi ,
\chi$ being the so called negative energy spinor which comes into
play at the Compton scale \cite{BjDr-64}. Because of this property as shown in 
detail
\cite{BGSnc-01}, there is now a covariant derivative given by, in
units, $\hbar = c=1$,
\begin{equation}
	\frac{\partial \chi}{\partial x^\mu} \to [\frac{\partial}{\partial
		x^\mu} - n A^\mu]\chi\label{6De12}
\end{equation}
where
\begin{equation}
	A^\mu = \Gamma^{\mu \sigma}_{\sigma} = \frac{\partial} {\partial
		x^\mu} log (\sqrt{|g|})\label{6De13}
\end{equation}
$\Gamma$ denoting the Christofell symbols.\\
$A^\mu$ in (\ref{6De13}) is now identified with the electromagnetic
potential, exactly as in Weyl's theory except that now, $A^\mu$
arises from the bi spinorial
character of the Dirac wave function.
Thus the desired unification is achieved.\\
Finally it may be added  that this is already in the domain 
of the \textit{Theory Of Everything} if the electroweak unificcation 
is taken into consideration.


\begin{thebibliography}{99}
\bibitem {harry} \emph{https://www.ted.com/talks/harry
	\_cliff\_have\_we\_reached\_the\_end\_of\_physics}
\bibitem {ijmpa1998} Sidharth, B.G. (1998). \emph{Int.J.Mod.Phys.A.} 13 (15), 1998, p.2599ff.
\bibitem {mg8} B.G. Sidharth, Proc. of the Eighth Marcell Grossmann Meeting on General Relativity, 1997,
Ed. T. Piran, World Scientific, Singapore, 1999, p.476-479.
 \bibitem {perl} S. Perlmutter, et al., Nature, Vol.391, 1 January 1998, p.51-59.
 \bibitem {leggett} Antony Leggett, private communication
 \bibitem {weinbergcc} Weinberg, S. (1989). \emph{The cosmological constant
problem} \emph{Reviews of Modern Physics}, Vol.6, No.1, January
1989, pp.1-23.
\bibitem {nernst} Kragh, H. (2012). \emph{Astronomy and Gravitation} (Royal Astronomical Society) Vol.53, February 2012.
\bibitem {rosen} Rosen, N. (1993). \emph{Int.J.Th.Phys.}, 32, (8), pp.1435--1440.
\bibitem {wheeler} Wheeler, J.A. and Kenneth, F. (1993).
\emph{Geons, Black Holes and Quantum Foam}, (W.W. Norton \& Co. New
York, 1993).
\bibitem {ijpap} Sidharth, B.G. (1997). \emph{Ind.J. of Pure and Applied Physics} 35, pp.456ff.
\bibitem {cu} Sidharth, B.G. (2001). \emph{Chaotic Universe: From the Planck to the Hubble Scale}
(Nova Science, New York).
\bibitem {feynman} Feynman, R.P. (1995). \emph{Feynman Lectures on Gravitation}
(Addison-Wesley Publishing Company, Reading, mass.).
\bibitem {uof} Sidharth, B.G. (2005). \emph{The Universe of Fluctuations} (Springer,
Netherlands).
\bibitem {bgs2003} Sidharth, B.G. (2003). \emph{The New Cosmos} in
\emph{Chaos, Solitons and Fractals} 18 (1) 2003, pp.197-201.
\bibitem {bgsnlw} Sidharth, B.G. (1994). \emph{Nonlinear World}1, pp.403--408.
\bibitem {bgsjsp} Sidharth, B.G. (1999). \emph{J.Stat.Phys.} 95, (3/4), pp.775--784.
\bibitem {beck} Beck, C. and Mackey, M.C. (2007). \emph{Electromagnetic Dark Energy} \emph{arXiv:astro-ph/0703364 v2} 27 August 2007.
\bibitem {wigner} Salecker, H. and Wigner, E.P. (1958). \emph{Quantum Limitations of the Measurement of
Space-Time Distances} \emph{Physical Review} Vol.109, No.2, January
15 1958, pp.571-577.
\bibitem {cer} Cercignani, C. (1998). \emph{Found.Phys.Lett.} Vol.11, No.2, 
pp.189-199.
\bibitem {BGSTU-08} Sidharth, B. G. (2008). The Thermodynamic Universe: 
Exploring the Limits of Physics. Singapore: World Scientific.

\bibitem{BjDr-64}Bjorken, J. D., Drell, S. D. (1964). Relativistic Quantum 
Mechanics. United Kingdom: McGraw-Hill.
\bibitem{BGSnc-01} Sidharth, B. G. (2001)\emph{Nuovo Cimento} Volume 116 B pp. 
735 ff.
\end{thebibliography}
\end{document}